\documentclass[]{interact}

\usepackage{epstopdf}
\usepackage{subfigure}
\usepackage{soul}
\usepackage{url}
\usepackage[hidelinks]{hyperref}
\usepackage[utf8]{inputenc}
\usepackage[small]{caption}
\usepackage{graphicx}
\usepackage{amsmath}
\usepackage{booktabs}
\usepackage{tabularx}
\usepackage{multirow}
\usepackage{times}
\usepackage{todonotes}
\usepackage{natbib}
\usepackage{float}
\usepackage{natbib}
\bibpunct[, ]{(}{)}{;}{a}{}{,}
\renewcommand\bibfont{\fontsize{10}{12}\selectfont}

\theoremstyle{plain}

\theoremstyle{definition}

\theoremstyle{remark}

\begin{document}


\title{Ascertaining price formation in cryptocurrency markets with Deep Learning}

\author{
\name{Fan~Fang\textsuperscript{a}\thanks{Contact Author. Email: fan.fang@kcl.ac.uk}, Waichung~Chung\textsuperscript{b}, Carmine~Ventre\textsuperscript{a}, Michail~Basios\textsuperscript{c,d}, Leslie~Kanthan\textsuperscript{c,d}, Lingbo~Li\textsuperscript{d}, Fan~Wu\textsuperscript{d}}
\affil{
\textsuperscript{a}King's College London, UK; 
\textsuperscript{b}University of Essex, UK; 
\textsuperscript{c}University College London, UK; \\
\textsuperscript{d}Turing Intelligence Technology Limited, UK}
}

\maketitle

\begin{abstract}
The cryptocurrency market is amongst the fastest-growing of all the financial markets in the world. 
Unlike traditional markets, such as equities, foreign exchange and commodities, cryptocurrency market is considered to have larger volatility and illiquidity. 
This paper is inspired by the recent success of using deep learning for stock market prediction. 
In this work, we analyze and present the characteristics of the cryptocurrency market in a high-frequency setting.  
In particular, we applied a deep learning approach to predict the direction of the mid-price changes on the upcoming tick.
We monitored live tick-level data from $8$ cryptocurrency pairs and applied both statistical and machine learning techniques to provide a live prediction. 
We reveal that promising results are possible for cryptocurrencies, and in particular, we achieve a consistent $78\%$ accuracy on the prediction of the mid-price movement on \emph{live} exchange rate of Bitcoins vs US dollars. 

\end{abstract}


\section{Introduction}

How algorithmic traders use impartial deep learning model and get an efficient prediction of price change is an important question for algorithmic trading. 
This paper focuses on effectively applying deep neural network on cryptocurrency market trading systems. Our objective is to predict the price changes; we consider both binary (up/down) and multi-class (e.g., degrees of increase/decrease) prediction of price changes. 

The cryptocurrency market is a huge emerging market~\citep{Aha13}. 
There were over $11,641$ exchanges available on the internet as of July 2018~\citep{Wik18}. 
Most of them are exchanges of small capitalisation with low liquidity. 
Exchanges with the highest 24-hour volume are FCoin, BitMEX, and Binance. 
Bitcoin, as the pioneer and also the market leader, has a market capitalization of over 112 billion USD, and a 24-hour volume over 3.8 billion USD in early July 2018. 
The cryptocurrency market is one of the most rapidly growing markets in the world, and are also considered one of the most volatile markets to trade in. 
For example, the price of a single Bitcoin increased significantly, from near zero in 2013 to nearly $19,000$ USD in 2017. For some alt-coins, the price can increase or fall over 50\% within a day. 
Therefore, having a 
method to accurately predict these changes is a pervasive task, but one that could achieve a long-term profit to cryptocurrency traders. 

There are a number of research papers that studied the structure of the limit order book (i.e., the bids at both sides of the market) and, more generally, the micro-structure of the market by using different methods ranging from stochastic to statistical and machine learning approaches~\citep{Hua05,Alt05,Fle12,Bia95}. 
The objective is to understand the way prices at either side of the market move; motivated by related literature we focus on a particular measure, the \emph{mid-price}, which intuitively captures the average difference between the best ask (the lowest price sellers are willing to accept) and best bid (the highest price buyers are willing to pay). 
Towards this aim we could, for example, use Markov chains to model the limit order book~\citep{Kel17}. 
We could view the limit order book as a queuing system with a random process and use birth-and-death chains to model its behaviour. 
From this perspective, a very intuitive way to explain the mid-price movement is to consider the value of the mid-price as the state of the chain. 
This value is controlled by the ratio between the probability of birth transitions $p$ and the probability of death transitions $q$. 
A ratio $p/q$ greater than 1 within a short interval of time indicates that there is a higher chance for birth transitions to happen (more buyers), and the value of the mid-price is expected to increase. 
Similarly, if the ratio is smaller than $1$, the value of the mid-price is expected to decrease~\citep{Sun09}.
The problem with this approach is the way it models the order book, namely, is the limit order book a queuing system? Even if it were, how to correctly simulate the random process and how to accurately estimate $p$ and $q$ become vital questions, with unclear answers, for this approach. 

In this research, we propose to 
adopt a deep learning approach to reveal useful patterns from the limit order book. We give answers to a number of very specific technical questions that have to do with this approach; our findings pave the way to the design of novel trading strategies and market estimators. A notable finding is that we can tweak the deep learning tools to achieve a consistent $78\%$ accuracy on the binary prediction of the movements of the mid-price on the live exchange of BTC-USD. 


\subsection{Related work}
Since the birth of the market, traders have been trying to find accurate models to use to make a profit. 
Many studies and experiments have been conducted based on statistical modelling of the stock price data. 
Some studies attempted to model the limit order book by using statistical approaches, such as using Poisson Processes and Hawkes Processes to estimate the next coming order and to model the state of the limit order book~\citep{Abe15,Mun11}. 

Others have used machine learning approaches to estimate the upcoming market condition by applying different machine learning models, such as support vector machine (SVM)~\citep{Ker15}, convolutional neural network (CNN)~\citep{Tsa17}, and recurrent network such as Long-Short-Term-Memory (LSTM)~\citep{Dix18}. These studies show that it is possible to use a data-driven approach to discover hidden patterns within the market. In particular, \citet{Ker15} modelled the high-frequency
limit order book dynamics by using SVM. They discovered that some of the
essential features of the order book lie on fundamental features, such as price and volume, and time-insensitive features like mid-price and bid-ask spread. 
In a more recent work, \citet{Sir18} also suggested that there might be some universal features on the stock market’s limit order book that have a non-linear relationship to the price change. They tried to predict the mid-price movement of the next tick by training a neural network using a significant amount of stock data. Their findings suggest that instead of building a stock-specific model, a universal model for all kinds of stock could be built. 

Most of the studies in the area focus on the traditional stock market like NYSE and NASDAQ~\citep{Gur11}.
Many researchers have studied these exchanges for many years. The quality of data and the
market environment are more desirable than those of the cryptocurrency market. 
Although the traditional stock market may provide a less volatile and more regulated environment for traders, the high volatility of the cryptocurrency market may provide a higher potential return. 
Our research aims to apply the same philosophy to the cryptocurrency market and replicate the findings above. 
In other words, we try to model the cryptocurrency market by using a data-driven approach. In this paper, experiments are focusing on the engineering side of this approach.

Understanding price formation through mid-price for cryptocurrencies becomes even harder due to the fact that they are distinct from traditional fiat-based currencies. 
The latter are usually issued by banks or governments. 
The only way to create Bitcoins, the currently dominant cryptocurrency, is to run a computationally intensive algorithm to add new blocks to the blockchain. 
People who participate in this processing will verify transactions on the blockchain, and try to earn Bitcoins as the reward of adding new blocks. 
These people are usually referred to as Bitcoins miners. The protocol of the Bitcoin fixed its total supply at 21 million~\citep{Nak08}.
Every transaction on the blockchain is protected by a cryptographic hash algorithm called SHA-256. 
It is a computational intensive hash algorithm that is implemented to verify blocks on the blockchain. For instance, if a counterfeiter wants to forge a block on the blockchain, they will also need to redo all the hashing before that block. 
This property provides a trustless foundation for Bitcoin because neither an individual nor an institution can counterfeit the currency or the transaction unless it has a computational power in excess of the majority of the network~\citep{Nak08}. 

Multi-label prediction is widely used in image processing, character recognition and forecasting of decisions or time series. Complex trading strategies might require more than binary classification. One may use the status box method to measure different stock statuses such as turning point, flat box and up-down box~\citep{Zha16} in order to reflect the relative position of the stock and classify whether the state coincides with the stock price trend.
In this research, we propose to use the transaction fee as a threshold to decide whether the designated cryptocurrency market has a long or short signal. When prediction of price movement is under transaction fee, it means in the next trading cycle the market falls in a 'Buffer area' where the market is in a relatively stable position. 

\subsection{Roadmap} 
The paper is organized as follows. In Section \ref{sec:dl}, we provide a brief overview of the tools adopted, including deep learning, limit order books and data sources we used. In Section \ref{sec:exp}, we design experiments to address our research questions. In Section \ref{sec:lim}, we give a brief discussion of the validity of our findings. Section \ref{sec:con} gives a conclusion of this paper.

\section{Our tools}\label{sec:dl}
In this section, we first review the background of deep learning and limit order books, before introducing an overview of the trading system where our prediction model is trained on. 

\subsection{Deep Learning}
Artificial neural networks are computational algorithms mimicking biological neural systems,
such as the human brain. 
These algorithms are designed to recognize and generalize patterns from the input, and remember them as weights in the neural network. 
The basic unit of a neural network is a neuron; a simple neural network, which is a conglomeration of neurons, is called Perceptron. 

The network used in this paper is a type of recurrent neural network called Long-Short-Term-Memory (LSTM)~\citep{Hoc97}. This is distinct from the feed-forward neural network such as Perceptrons, since the output of the neural network sends feedback to the input and affects the subsequent output. 
Therefore, LSTM is better suited for handling sequential data where the previous data can have an impact on subsequent data; this, in principle, works well for time series data for price prediction and forecasting.

\begin{figure}[H]
\centering
\includegraphics[width=0.7\columnwidth]{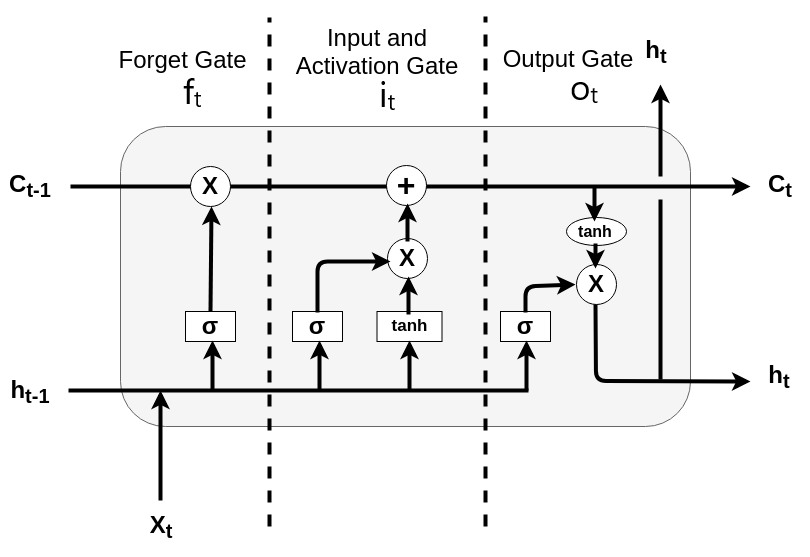}
\caption{An overview of an LSTM cell}
\end{figure}
An LSTM cell contains a few gates and a cell status to help the LSTM cell decide what
information should be kept and what information should be forgotten. As a result, the LSTM cell
can recall important features from the previous prediction by having a cell state. An LSTM cell
can also be viewed as a combination of a few simple neural networks, each of them serving a
different purpose.
The first one is the forget gate~\citep{Hoc97}. The previous output is concatenated with the new
input and passed through a sigmoid function. After that, the output of the forget gate, $f_t$, will
perform a Hadamard product (element-wise product) with the previous cell’s state. Note that $f_t$
is a vector containing elements that have a range from 0 to 1. A number closer to 0 means the
LSTM should not recall it, whilst a number closer to 1 means the LSTM should recall and carry on to
the next operation. This process helps the LSTM select which elements are to forget and
remember, respectively.
The second one is the input and activation gates~\citep{Hoc97}. This process concatenates the
previous output with the new input, determines which element should be ignored, and updates
the internal cell state. 
The cell state is then updated by a combination of the output and a transformation of the input.
The third one is the output gate~\citep{Hoc97}. This process helps determine the output of the cell.
Finally, the output of the LSTM cell is the Hadamard product of the current internal cell state and the output of the output gate~\citep{lstmmath,lstmexplain}. 

We use Root Mean Square Propagation (RMSprop)~\citep{Tie12} -- a 
stochastic gradient descent optimizer -- to train the neural network, with the learning rate divided by the exponentially weighted average. 
Optimizer, learning rate and loss function are core concepts in deep learning models. Optimizer ties together the loss function and model parameters by updating the model in response to the output of the loss function. Loss function is a method of evaluating how well your algorithm models your dataset, which tells the optimizer when it’s moving in the right or wrong direction. The learning rate is a hyperparameter that controls how much to change the model in response to the estimated error each time the model's weights are updated.

In our experiments, we also tested the use of an adaptive moment estimation, Adam in short,
as the optimizer. While we observed that Adam helps the neural network to converge faster, we noted a tendency to overfit the data; the validation set has an increasing loss while the training set has a decreasing loss. This motivates our choice of RMSprop as optimizer.

\subsection{Limit Order Books}
The limit order book is technically a log file in the exchange showing the queue of the outstanding
orders based on their price and arrival time. Let $p_b$ be the highest price at the buy side,
which is called the best bid. The best bid is the highest price that a trader is willing to pay to
buy the asset. Let $p_a$ be the lowest price at the sell side, which is called the best ask. The best
ask is the lowest price a trader is willing to accept for selling the asset. 

The mid-price of an asset is the average of the best bid and the best ask of the asset in the market.
\begin{equation*}
M_p = \frac{(p_b + p_a)}{2}.
\end{equation*}
There are other metrics that are also useful
for describing the state of the limit order book: Spread, Depth and Slope.

\subsection{Data source and  overview of the envisioned trading system}
Numerous exchanges provide Application Programming Interface (API) for systematic traders
or algorithmic traders to connect to the exchange via software. Usually, an exchange provides two types of API, a RESTful API, and WebSocket API. Some exchanges also provide a Financial Information eXchange (FIX) protocol. In this study, a WebSocket API from an exchange called GDAX (Global Digital Asset Exchange) is used to retrieve the
level-2 limit order book live data~\citep{GDA18}. The level-2 data provides prices and aggregated depths for top 50 bids and asks. GDAX is one of the largest exchanges in the world owned by the Coinbase company.

\begin{figure}[h]
\centering
\includegraphics[width=0.7\columnwidth]{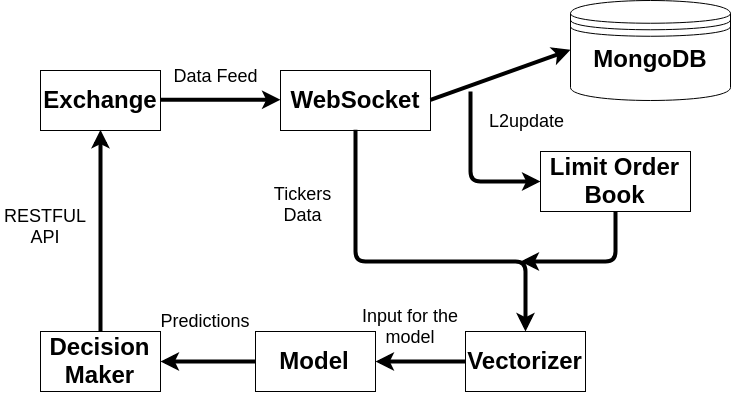}
\caption{An overview of a simple trading system}
\label{fig:tradingsystem}
\end{figure}

Our focus is to design a model that can successfully predict the mid-price movement in the context of cryptocurrencies. Such a model is a component of a trading system, as shown in Figure \ref{fig:tradingsystem}. There are a few essential components for the trading system. First of all, the WebSocket is used to subscribe to the exchange and receive live data including tickers, order flows, and the limit order book’s update. Tickers data usually appears when two orders of the opposite side are matched and the opening of a candle on a candlestick chart. Tickers contain the best bid, best ask, and the price, thus reflecting the change in price in real-time.  


The way the updates to the limit order book are communicated differ. Some exchanges provide a real-time snapshot of the order book. Some exchanges, including GDAX, only provide the update, i.e., updated data of a specific price and volume on the limit order book. Therefore, a local real-time limit order book is required to synchronize with the exchange limit order book. Additionally, we need to store all the data in a database. In this study, a non-relational database called MongoDB has been used to this purpose. Unlike a traditional relational database, MongoDB stores unstructured data in a JSON-like format as a collection of documents. The advantage of using a non-relational database is that data can be stored in a more flexible way. 
The local copy of the limit order book is reconstructed by using level-2 limit order book updates. The reconstructed limit order can provide information on the shape and status of the exchange limit order book. This limit order book can be used for calculating order imbalance and can provide quantified features of the limit order book. The input to the model is then finalized by a vectorizer, used as a data parser, combining information and extracting features from the ticker data and the local limit order book. Features are then reshaped into the format that can fit into the trained LSTM model.

We leave to future research the design and experimentation of a decision maker, which should make use of the prediction given by the trained model and help manage the inventory. If the inventory and certain thresholds are met, the decision-maker would place an order to the exchange based on the prediction from the trained LSTM model through RESTful API. 

\section{Experimental Study}\label{sec:exp}
\subsection{Objective}

The purpose of this research is to process real-time tick data using deep learning neural network approach on cryptocurrency trading system. As a deep learning model based on high-frequency trading, accuracy of prediction and computational efficiency are both important factors to consider in this research. 

\subsection{Dataset}

The data used in this study is live data recorded via a WebSocket through the GDAX exchange
WebSocket API. The data contain the ticker data, level-2 order book updates, and the
order submitted to the exchange. The time range of the collected data is from the time of
2018-07-02T17:22:14.812000Z to 2018-07-03T23:32:53.515000Z. The order flow data contain
$61,909,286$ records, the tickers data include $128,593$ ticker data points, and the level-2 data
contain $40,951,846$ records. Table 1 lists the available assets on the GDAX exchange and the
corresponding number of records.

\begin{table}[h]
    \caption{Amount of data collected}
    \begin{tabularx}{\columnwidth}{cXXX}
        \hline
        product id \textbackslash data type & Ticker & Level-2 & Order Flow    \\
        \hline
		BCH-USD & 15,213 & 1,600,474 & 2,442,323 \\
		BTC-EUR & 9,769 & 4,656,627 & 7,002,588 \\
		BTC-GBP & 3,726 & 8,849,556 &13,280,280\\
		BTC-USD & 25,904 & 4,110,818 & 6,282,022 \\
		ETH-BTC & 4,016 & 1,250,202 & 1,893,851 \\
		ETH-EUR & 3,180 & 4,876,886 & 7,323,178 \\
		ETH-USD & 27,089 & 6,087,574 & 9,276,806 \\
		LTC-BTC & 2,167 & 611,682 & 923,070 \\
		LTC-EUR & 4,243 & 1,260,024 & 1,897,731 \\
		LTC-USD & 32,203 & 2,391,377 & 3,700,271 \\
		BCH-EUR & 4,243 & 5,822,103 & 7,934,653 \\
        \hline
    \end{tabularx}
\end{table}

\subsection{Methodology}

\subsubsection{Model Architecture}

\begin{figure}[h]
\centering
\includegraphics[width=0.8\columnwidth]{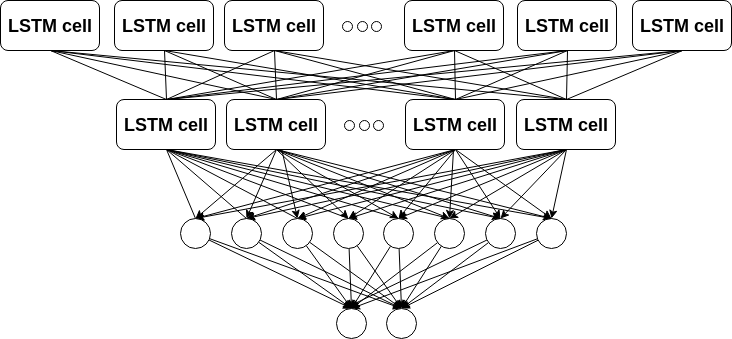}
\caption{LSTM model architecture}
\label{fig:architecture}
\end{figure}

The simple architecture in Figure \ref{fig:architecture} served as the predictive model in this study. This neural network
contains two layers of LSTM cells, one layer of fully connected neurons, and one layer of softmax
as the output layer which outputs the probability of going up or going down. The two layers of LSTM cells can be viewed as a filter for capturing non-linear features from the data, and the fully connected layer can be viewed
as the decision layer based on the features provided by the last LSTM layer. This neural network
is designed as simple as possible because in the tick data environment, every millisecond
matters. Reducing the number of layers and neurons can significantly reduce the computational
complexity, thus the time required for the data processing.

\subsubsection{Multi-label prediction}
Binary classification can be scarcely informative to a trader, as ``small'' variations are not differentiated from ``big'' ones. One might want to hold one's position in the former case and transact only in the latter.

We use 1-min and 5-min data to demonstrate the rate of price change, defined as the ratio between the price change and the transaction (close) price. In both cases, most relative price changes fall in $-0.25\%$ and $0.25\%$. Often these percentages are less than the transaction fees and traders ought to be able to know when this is the case to develop a successful trading strategy. Therefore, we also investigate multi-label prediction based on trading strategy needs. 
In this multi-label prediction, we replace binary target prediction with four-target prediction. At the structure level, we have four softmax units as output layer instead of two units. By effectively set the boundaries of four units, we can transform the original two-class classifier into a four-class classifier. 
Using the fees used by Coinbase Pro~\citep{Cob18}, we use $\pm0.2\%$ of the transaction price as a reasonable threshold to differentiate large and small changes, see Table~\ref{tbl:multilab} where we also name the intervals for convenience.

\begin{figure}[h]
\centering
\includegraphics[width=0.7\columnwidth]{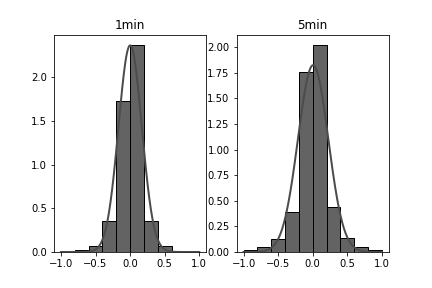}
\caption{Distribution of historical price changes}
\end{figure}

\begin{table}
\centering
\caption{Multi-label prediction}
\begin{tabular}{lcl}
\hline
\multicolumn{1}{c}{Label}                    & \multicolumn{1}{c}{Relative Price Change}       & \multicolumn{1}{c}{Type}                                  \\ \hline
Significant increase   & $(+0.2\%,+ \infty)$ & \multirow{2}{*}{Sensitive Interval}   \\
Significant decrease   & $(-0.2\%,- \infty)$ &                                       \\
Insignificant increase & $(0,+0.2\%]$  & \multirow{2}{*}{Insensitive Interval} \\
Insignificant decrease & $[-0.2\%,0)$  &                                       \\ \hline
\end{tabular}
\label{tbl:multilab}
\end{table}

\subsubsection{Walkthrough Training}
Prediction model in financial market has timeliness; this is especially true for the high-frequency financial market. 
For example, should we use historical financial data from 2015 to train a model and test it on 2017 data for predictions, this model might not have a good performance. The old model might not adapt well to the new market environment as it has been trained and tailored on old market conditions. Although a deep learning approach can largely increase prediction accuracy of stock market, such models need to optimize themselves because the stock market is constantly changing.

Sheng et al.~\citep{wan2006parameter} propose the parameter incremental learning (PIL) method for neural networks; the main idea is that the learning algorithm should not only adapt to the newly presented input-output training pattern by adjusting parameters, but also preserve the prior results.
Inspired from this, we propose a method called \emph{Walkthrough Training} in deep learning for our task. This approach is designed to retrain the original deep learning model itself when it ``appears'' to no longer be valid. We consider two different Walkthrough training methods. 
\begin{itemize}
	\item[(i).] \emph{Walkthrough with stable retrain frequency.} Considering different trading cycles based on the data obtained from the API, we retrain our model at fixed time intervals. The length of the interval depends on our trading strategy and accuracy from data we obtained. This way of retraining helps the model to adjust to the newly acquired features and retain the knowledge gained from the original training.
	\item[(ii).]  \emph{Walkthough with dynamic retrain frequency.}
	We use Maximum Drawdown (MDD), which is the maximum observed loss from a peak to a trough of a portfolio before a new peak is attained, as a condition of dynamic retraining. 
	The idea is that stable retraining is not suitable for every condition in retraining model. More specifically, if the old model is aimed to long-term prediction, stable retraining will lead to waste of computing resources and overfitting problem (the model fits the data too well and leads to low prediction accuracy on unseen data). 


During the process of prediction based on this method, we monitor accuracy of prediction over time. In the following formula, "Min Accuracy Value" and ``Max Accuracy Value'' identify the highest and lowest prediction accuracy, respectively. All parameters in the formula are in interval between last retraining time and current calculation time. After calculation, ``Modified MDD'' is considered as hyper-parameter in the whole prediction model to optimize the retraining time. 
\begin{equation*}
\text{Modified MDD} = \frac{\text{Max Accuracy Value - Min Accuracy Value}}{\text{Min accuracy Value}}. \end{equation*}
The modified MDD is a measure of drawdown that looks for greatest effective period of model. When modified MDD is over 15\%, we consider the original deep learning model to be no longer applicable for latest market data. In such a case, we use historical data up to the point when the MDD is measured as training data to retrain original deep learning model. This process will be through whole time series prediction. 
\end{itemize}

\subsection{Research Questions}
We investigate four specific research questions (RQs, for short) in our general context of interest, price predictions through a deep learning model within the cryptocurrency markets.

\begin{description}
	\item[RQ1:] {\bf How well does a universal deep learning model perform?}\\ \citet{Sir18} found that a universal deep learning model would predict well the price formation
	in relation to stock market. We ask this question to understand if a similar conclusion can be drawn for more emergent, less mature and more volatile cryptocurrency market.
	\item[RQ2:] {\bf How many successive data points should we use to train deep learning models?} \\The sequential nature of time series naturally puts forward the question of optimizing the number of subsequent data points (i.e., time steps) used to train the deep network. Does it make sense to use more than one data point at a time? If so, how many time steps should be used?
	\item[RQ3:] {\bf How well do deep learning models work on live data?} \\A good offline prediction based on deep learning may fail to perform well on live data, due to evolving patterns in a highly volatile environment like ours. Is there an accuracy decay on live data? If yes, would Walkthrough training methods help address the issue? Moreover, we want to understand if lean and fast architectures can perform well with tick online data. 
	\item[RQ4:] {\bf What is the best Walkthrough method in the context of multi-label prediction?}\\
	Making profit on tick data predictions might be too hard for a number of reasons. Firstly, the execution time of the order might make the prediction on the next tick obsolete. Secondly, in the context of multi-label predictions, there might be very few data points in the sensitive intervals which would make transactions potentially more profitable than transaction costs. We therefore wish to determine the best Walkthrough method when we use minute-level data for the task of multi-label classification. 
\end{description}
Ultimately, the findings from the questions above will help a cryptocurrency trader to design a better model and ultimately devise a more profitable trading strategy (i.e., the decision maker in the system of Figure \ref{fig:tradingsystem}).

\subsection{Results and Analysis}
We organize the discussion of our results according to the research questions of interest. The answer to each question informs the design used to address the challenges of the subsequent questions. In this sense, we use an incremental approach to find our results. 


\subsubsection{Answer to RQ1: How well does a universal deep learning model perform?}
We begin by training product specific networks of Figure \ref{fig:architecture} in order to establish the baseline for comparison. For each product (i.e., currency pair), five neural networks having the same architecture are initialized. Five training sets are then created by extracting the first 10\%, 20\%, 50\%, 70\%, and 85\% from the total data of the product. After that, the neural network is trained and tested with each data split. For example, a product-specific, such as BCH-USD, neural network is trained with the first 10\% of the total data using only one time step; the rest of the data are then used to evaluate the performance of the neural network. Subsequently, another neural network is trained and tested with a different amount of data and so on.

The purpose of using this training approach is to evaluate the importance of the amount of
data used. The high-frequency markets are often considered extremely
noisy and full of unpredictability. If neural networks for the same product showed no performance gain
with increasing amount of training data, then it may actually be the case that the majority of the data
is actually noise. In these circumstances, a stochastic model might be a better option than
 a data-driven model, because a simpler model generally tends to be less overfitting compared
to a complex model under noisy environment.

\begin{table}[H]
    \caption{Out-of-sample accuracy with respect to training sample sizes}

    \begin{tabularx}{\columnwidth}{cXXXXX}
    	\hline
    	currency pair & \multicolumn{5}{c}{Sample size used in training} \\ 
        \hline
         & 10\% & 20\% & 50\% & 70\% & 85\%  \\
        \hline
		BCH-USD &  0.619 & 0.664 & 0.674 & 0.699 & 0.662 \\
		BTC-EUR &  0.554 & 0.561 & 0.551 & 0.596 & 0.470 \\
		BTC-GBP &  0.611 & 0.551 & 0.634 & 0.565 & 0.540 \\
		BTC-USD &  0.702 & 0.789 & 0.797 & 0.825 & 0.814 \\
		ETH-BTC &  0.788 & 0.825 & 0.839 & 0.775 & 0.743 \\
		ETH-EUR &  0.633 & 0.640 & 0.714 & 0.658 & 0.597 \\
		ETH-USD &  0.599 & 0.608 & 0.555 & 0.703 & 0.736 \\
		LTC-BTC &  0.579 & 0.687 & 0.738 & 0.751 & 0.730 \\
		LTC-EUR &  0.505 & 0.503 & 0.586 & 0.602 & 0.672 \\
		LTC-USD &  0.574 & 0.620 & 0.767 & 0.787 & 0.814 \\
		BCH-EUR &  0.540 & 0.540 & 0.540 & 0.432 & 0.526 \\
        \hline
    \end{tabularx}
\label{tbl:trainingset}
\end{table}

\begin{figure}[H]
\centering
\includegraphics[width=0.8\columnwidth]{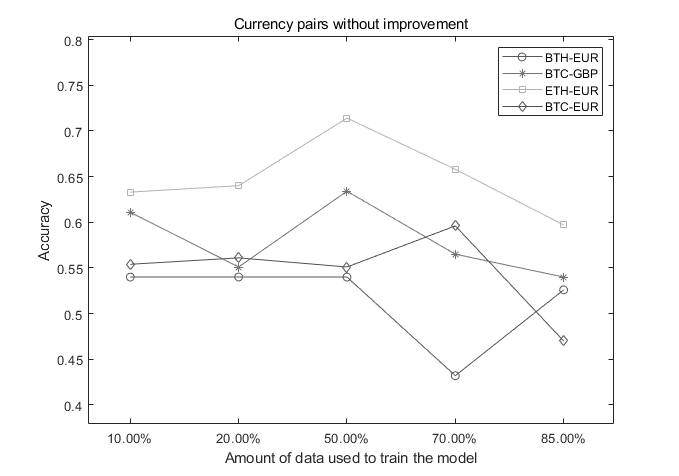}
\caption{Currency pairs without improvement}
\label{fig:noimp}
\end{figure}

\begin{figure}[H]
\centering
\includegraphics[width=0.8\columnwidth]{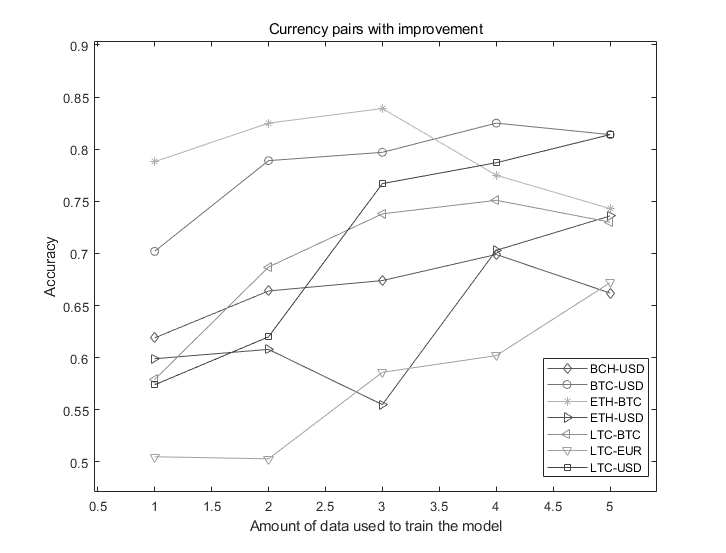}
\caption{Currency pairs with improvement}
\label{fig:imp}
\end{figure}

\begin{figure}[H]
\centering
\includegraphics[width=0.8\columnwidth]{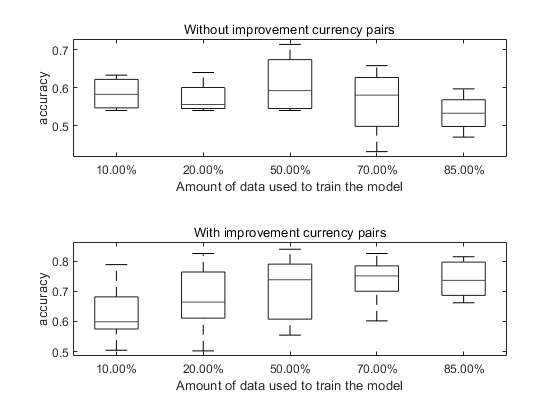}
\caption{Box plots of currency pair with and without improvement}
\label{fig:boxp:pairs}
\end{figure}
From the result in Table~\ref{tbl:trainingset}, the currency pairs with very little samples, such as BCH-EUR, BTC-GBP, ETH-EUR, and BTC-EUR, show a decreasing performance after using training data with a size greater than 50\% (shown as Figure~\ref{fig:noimp}). The decrease in the performance could be a direct result of the lack of testing cases. For other currency pairs, the currency-pair-specified neural network models show a general rise in accuracy when increasing the size of the training data (Figure~\ref{fig:imp}), which suggests that there might be some recognizable patterns in the data. The box plots (Figure~\ref{fig:boxp:pairs}) show the comparison of currency pairs with and without improvement. The result above suggests that, at least for our architecture, the neural network is able to learn the hidden pattern from within a dataset when given a sufficient amount of data for most of the currency pairs. 

We are now ready to test the findings of \citet{Sir18} about the existence of a universal predictive model in the context of cryptocurrencies. We are interested to see whether a universal predictive model for all available currency pairs can outperform the product-specific ones introduced above. Table~\ref{tbl:univ} displays the performance of different models. The label ``AVG'' represents the mean performance of all models. We know from the analysis above that for some currency pairs, the current neural network architecture is not performing very well. Therefore, for more precise and targeted analysis, those currency pairs are excluded from the original dataset, and a new dataset is generated without them. The label of ``selected'' represents the mean performance of all models excluding those pairs, namely, BCH-EUR, BTC-GBP, ETH-EUR and BTC-EUR. The ``universal selected'' neural network is trained with the same approach but with joined data across all available products. 

\begin{table}[H]
\centering
\caption{Models' performance with different sample sizes used in training}
\begin{tabular}{@{}ccccc@{}}
\toprule
            & \multicolumn{4}{c}{Model accuracy on testing set}                                             \\ \midrule
Sample size & AVG     & Selected & Universal & \begin{tabular}[c]{@{}c@{}}Universal\\ Selected\end{tabular} \\ \midrule
10\%         & 0.60990 & 0.62419  & 0.64899   & 0.66417                                                      \\
20\%         & 0.63578 & 0.67134  & 0.66962   & 0.68286                                                      \\
50\%         & 0.67263 & 0.70823  & 0.69180   & 0.70057                                                      \\
70\%         & 0.67252 & 0.73500  & 0.71111   & 0.73126                                                      \\
85\%         & 0.66429 & 0.73902  & 0.71533   & 0.74140                                                      \\ \bottomrule
\end{tabular}
\label{tbl:univ}
\end{table}

We can see that the universal model slightly outperforms the mean of product-specific
models, for each size of the training set, by an average of 3.65\% in terms of accuracy. Similarly, the universal with selected currency pairs outperforms the selected product-specific model by an average of 4.25\%. In general, both of the universal models achieved higher accuracy than the product-specific ones. Therefore, we can conclude that the universal model has better performance than the currency-pair specific model. The performance gain in the universal model and the universal model with selected currency pairs may be explained with the following rationale. Firstly, there are some universal features on the limit order book which could be observed by the LSTM neural network for most of the currency pairs on the exchange. Secondly, the increased amount of the training data helps the network to
generalize better, since 10\% of joined data is much larger than 10\% of one currency pair data.
It also means that the LSTM model can learn the pattern from the data of multiple currency pairs having the same time horizon, then apply the pattern to another currency pair.

We reach the following conclusion from this section. \textbf{The answer to RQ1} is that the universal model has better performance than the currency-pair specific model for all the available currency pairs in (the chosen) cryptocurrency market. 

\subsubsection{Answer to RQ2: How many successive data points should we use to train deep learning models?}
Informed by our findings in relation to RQ1, we next fix the training set size to 70\% of the total sample size and focus our attention to the universal and universal selected neural networks. To investigate RQ2, we train both networks with 70\% of the total data using increasing time steps of 1, 3, 5, 7, 10, 20, 40. For example, the 3-time-steps input contains the feature vector of the current tick $F_t$, feature vector of the previous tick $F_{t-1}$, and feature vector of two ticks prior $F_{t-2}$. This approach aims to discover whether there are any observable patterns related to the sequence of data and how persistent it is. 

\begin{table}[hbt]
\centering
\caption{Performance of the models for different time steps}
\begin{tabularx}{0.85\columnwidth}{XXX}
\hline
Time steps & Universal & Universal Selected \\ \hline
1          & 0.7111    & 0.7312             \\ 
3          & 0.7263    & 0.7445             \\
5          & 0.7200    & 0.7419             \\ 
7          & 0.7086    & 0.7369             \\ 
10         & 0.7146    & 0.7389             \\ 
20         & 0.7116    & 0.7421             \\ 
40         & 0.7131    & 0.7275             \\ \hline
\end{tabularx}
\label{tbl:steps}
\end{table}

The results are shown in Table \ref{tbl:steps}. To make sense of them, we fit the data points in a linear regression model, with ordinary least squares, and obtain the coefficients in Table \ref{tbl:ols}, where Const represents the intercept and X the slope.

\begin{table}[h]
\centering
\caption{Coefficients of model accuracy in OLS regression}
\begin{tabular}{@{}lrr@{}}
\toprule
      & \multicolumn{1}{l}{Universal} & \multicolumn{1}{l}{Universal Selected} \\ \midrule
Const & 0.7166                        & 0.7406                                 \\
X     & -0.0001                       & -0.0002                                \\ \bottomrule
\end{tabular}
\label{tbl:ols}
\end{table}

\begin{figure}[h]
\centering
\includegraphics[width=1\columnwidth]{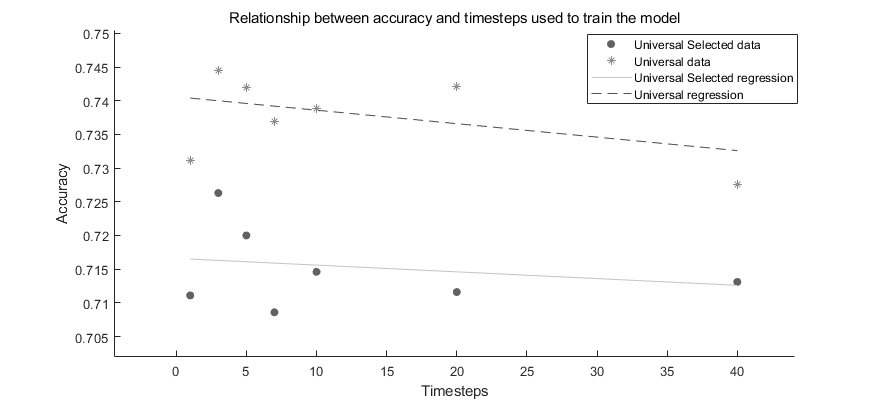}
\caption{Relationship between accuracy and number of time steps used in training}
\label{fig:steps}
\end{figure}

As depicted in Figure \ref{fig:steps}, the slopes of the linear equations are very close to zero, and are negative. This result suggests that increasing the time steps of the training data does not
have a significant effect on the performance of the model. On the contrary, increasing the time steps too much may also have a negative impact to the model’s performance. 

\textbf{The answer to RQ2} is that one time step/data point is the best choice in our context. Choosing one time step carries some further advantages; one step, in fact, opens the possibility of using different machine learning algorithms since most of them are not designed to handle sequential input. 

\subsubsection{Answer to RQ3: How well do deep learning models work on live data?}
In this section, the challenges and performances of using our predictive models on live data are discussed.

\begin{figure}[h]
\centering
\includegraphics[width=1\columnwidth]{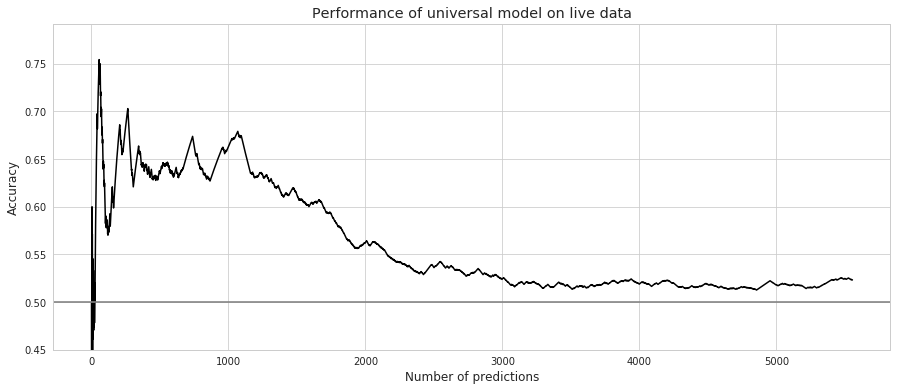}
\caption{Performance decay on the the live data}
\label{fig:decay}
\end{figure}

The 
horizontal line in Figure \ref{fig:decay} is the baseline of the performance, and the 
black line is the performance of the predictive model on live data. This figure shows that the performance of the universal model slowly decays to almost random guessing over the period of interest. This behavior could be caused by some non-stationary features of the limit order book, which means that the hidden pattern captured by the universal model is no longer applicable to
the new data.

\begin{figure}[h]
\centering
\includegraphics[width=0.6\columnwidth]{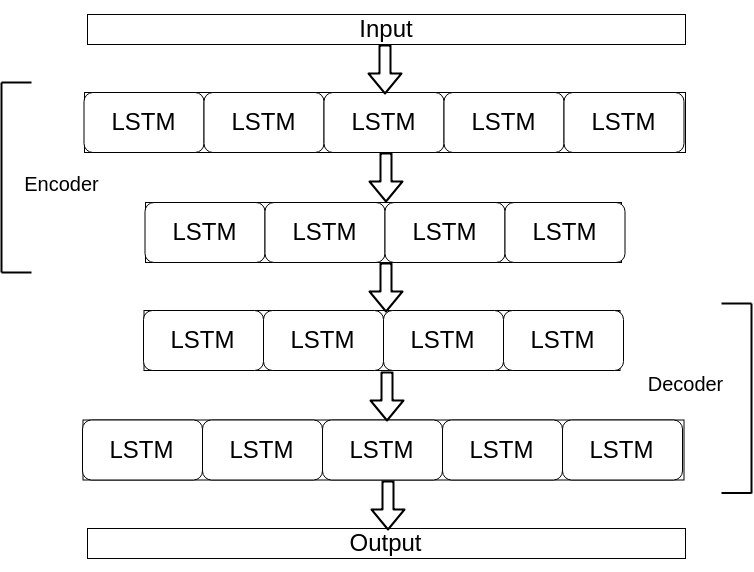}
\caption{Achitecture of the autoencoder}
\label{fig:autoenc}
\end{figure}

To resolve this problem, an autoencoder is used (Figure \ref{fig:autoenc}). The characteristic of
the autoencoder is that the input layer and the output layer usually have the same number of neurons, and the hidden layers of the autoencoder must have a lower number of neurons compared to the input and output layers. The reason for using such an architecture is that the reduced number of neurons in the hidden layers can form a bottleneck in the neural network. Thus, the autoencoder cannot learn by simply remembering the input only. This architecture, in fact, forces the autoencoder to compress the input data and then decompress the data before outputting it. Therefore, the autoencoder can learn from the input structure. The trained autoencoder performs two tasks. The first one is to remove noise; the trained autoencoder can suppress abnormal features by reconstructing the input data. This process usually removes abnormal spikes in a feature. The second one is to map the new data into a more familiar space for the LSTM model.

\begin{figure}[h]
\centering
\includegraphics[width=1\columnwidth]{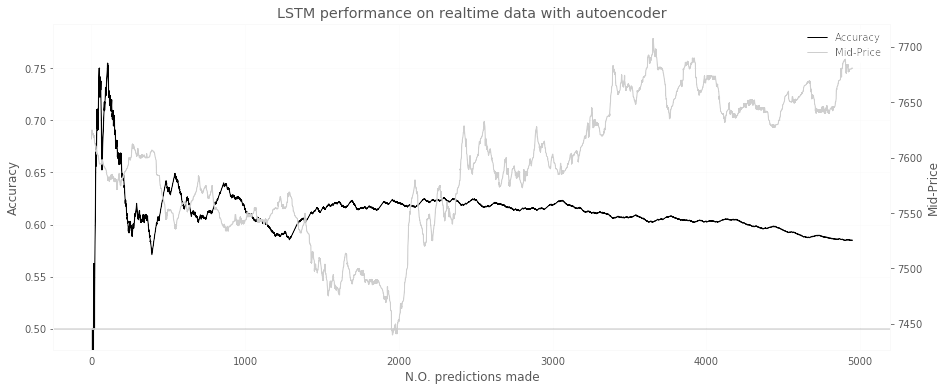}
\caption{Performance of the universal model with autoencoder}
\label{fig:autoencperf}
\end{figure}

\begin{figure}[h]
\centering
\includegraphics[width=0.8\columnwidth]{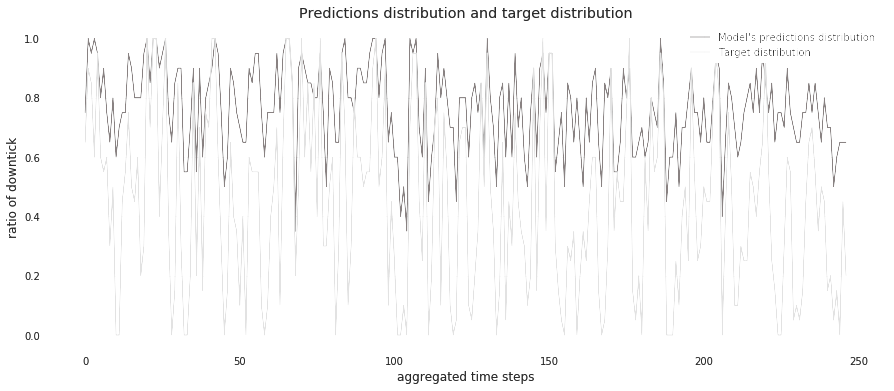}
\caption{Predictions distribution and real-time target distribution}
\label{fig:realdist}
\end{figure}

Figure \ref{fig:autoencperf} shows the prediction of the LSTM with an autoencoder by using live data of BTC-USD from 2018-08-01 15:10:43.674 to 2018-08-02 08:33:50.367. Bitcoin has a great dominance and the BTC-USD is also the most traded product on the market. The performance decays slower with the autoencoder than the original LSTM model. Figure \ref{fig:realdist} is the distribution of the
predictions made by the universal model with autoencoder and the aggregated real-time target; each point of the aggregated real-time target is equal to the mean of upticks and downticks for every 20 samples. 
The darker 
line depicts the ratio of downticks given by the predictive model, and the lighter 
line is the ratio of downticks given by real-time target. From the distribution of prediction and real-time target we can observe that the autoencoder is slightly biased to the downtrend market. This explains the gradual decrease in the accuracy under the uptrend market after the 3,000 predictions mark (cf. Figure \ref{fig:autoencperf}) because small errors accumulate over time and eventually affect the overall accuracy. In other words, the biased training data could cause a biased model. For example, the training data used to train the model could be experiencing a bearish market so that the model is more sensitive to the downtrends.

\begin{figure}[H]
\centering
\includegraphics[width=1\columnwidth]{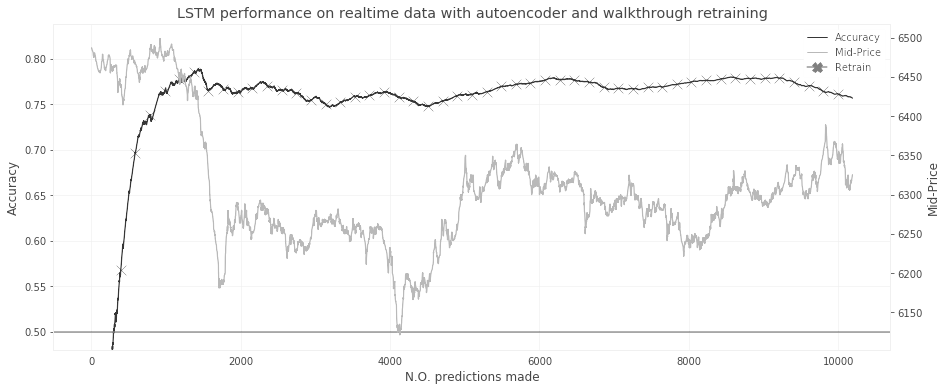}
\caption{Performance of the universal model with autoencoder}
\label{fig:btclive}
\end{figure}

An intuitive way to adjust the bias of the model is walkthrough training, so to retrain the model with recent data. This way, the model can learn from the most recent data, and integrate it with the original data. We implement a walkthrough with stable retrain time as follows. First, a queue buffer is set up to collect features from the live data. After every 196 predictions made by the model, the model retrains by the newly collected features in the buffer.

\begin{figure}[h]
\centering
\includegraphics[width=1\columnwidth]{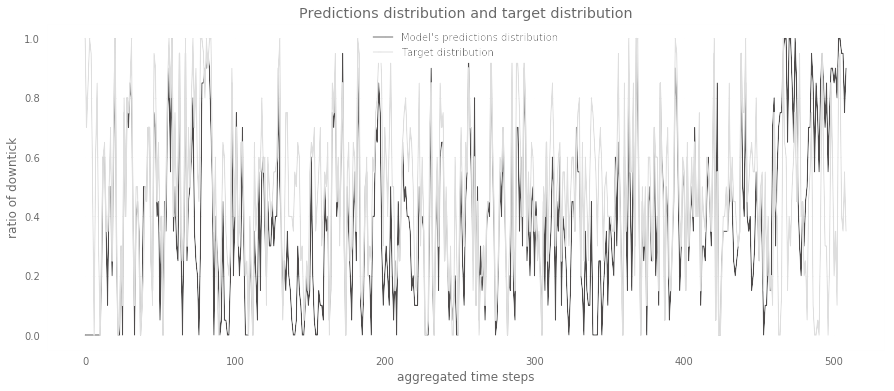}
\caption{Predictions distribution and real-time target distribution}
\label{fig:btcliverealdist}
\end{figure}

To test the effectiveness of this modification, we use live data of BTC-USD from 2018-08-08 14:31:54.664 to 2018-08-09 09:01:13.188. The results are plotted in Figure \ref{fig:btclive}. We observe that before the first retraining, the model lacks the predictive power on live data. It starts with an accuracy of less than 50\%, which is worse than random guessing. After the first few instances of retrain, however, the model improves accuracy from 58\% to 78\%, to finally stabilize around 76\%. Moreover, the distribution of the predictions of the model shows a similar shape to the real-time target distribution, and no apparent bias can be observed, cf. Figure \ref{fig:btcliverealdist}.

\begin{figure}[h]
\centering
\includegraphics[width=0.6\columnwidth]{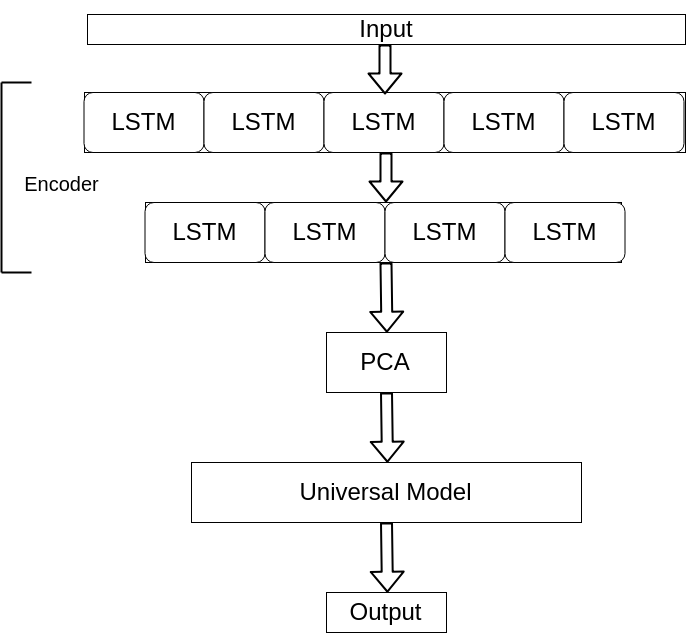}
\caption{LSTM model with autoencoder}
\label{fig:dimreduct}
\end{figure}

A further improvement of the model to work on live data is needed to improve the execution speed and reduce the chance of overfitting. This is achieved by reducing the dimension of the input data. The intermediate output of the autoencoder, which is the output of the encoder part, is used instead of using the original data. 
Because of the architecture of the autoencoder, the hidden layer contains fewer neurons than the output layer. Although the hidden layer contains fewer neurons, it preserves all the essential information of the input data. By using this approach, the universal model can use fewer neurons to capture the information that is needed to make predictions. 
Therefore, the neural network has less freedom to be overfitted, and the reduction of the size of the neural network also improves the execution speed.
Our architecture uses the intermediate encoder output as the input for the LSTM model, cf. Figure \ref{fig:dimreduct}. The advantage of using the autoencoder instead of using Principal Component Analysis (PCA) directly is that autoencoder
can map the 3D sequential data (sample size, time steps, features) into a vector. This process helps to capture the information from the sequence which could not be done by the PCA only (PCA can only deal with 2D data).

\textbf{The answer to RQ3} is summarized in Table \ref{tbl:liveresults}, where we display the performance metrics of the predictive model with a reduced architecture on the same live data of BTC-USD from 2018-08-08 14:31:54.664 to 2018-08-09 09:01:13.188. In the table
``autoencoder as denoiser''  refers to architecture in Figure \ref{fig:autoenc} (including encoder, decoder, PCA, universal model and output) whilst ``autoencoder as reducer'' refers to the architecture in Figure \ref{fig:dimreduct} (including encoder, PCA, universal model and output).
The difference of the precision score and the accuracy score is much lower than the original model, which suggests that there is less overfitting. Interestingly, the reduced model has a $2.43\%$ increase in the accuracy score. 

\begin{table}[h]
\centering
\caption{Classification report}
\begin{tabular}{lrrrr}
\hline
\multicolumn{5}{c}{Classification report of autoencoder as denoiser}                                                                                         \\ \hline
\multicolumn{1}{l|}{}            & \multicolumn{1}{l}{Precision} & \multicolumn{1}{l}{Accuracy} & \multicolumn{1}{l}{F1-score} & \multicolumn{1}{l}{Support} \\ \hline
\multicolumn{1}{l|}{up}          & 0.72                          & 0.86                         & 0.78                         & 5,265                       \\
\multicolumn{1}{l|}{down}        & 0.81                          & 0.65                         & 0.72                         & 4,927                       \\
\multicolumn{1}{l|}{}            &                               &                              &                              &                             \\
\multicolumn{1}{l|}{avg / total} & 0.77                          & 0.76                         & 0.75                         & 10,192                      \\ \hline
\multicolumn{5}{c}{Classification report of autoencoder as reducer}                                                                                          \\ \hline
\multicolumn{1}{l|}{}            & \multicolumn{1}{l}{Precision} & \multicolumn{1}{l}{Accuracy} & \multicolumn{1}{l}{F1-score} & \multicolumn{1}{l}{Support} \\ \hline
\multicolumn{1}{l|}{up}          & 0.79                          & 0.78                         & 0.79                         & 5,265                       \\
\multicolumn{1}{l|}{down}        & 0.77                          & 0.78                         & 0.77                         & 4,927                       \\
\multicolumn{1}{l|}{}            &                               &                              &                              &                             \\
\multicolumn{1}{l|}{avg / total} & 0.78                          & 0.78                         & 0.78                         & 10,192                      \\ \hline
\end{tabular}
\label{tbl:liveresults}
\end{table}

\subsubsection{Answer to RQ4: What is the best Walkthrough method in the context of multi-label prediction?}
In this section, we concentrate on multi-label prediction using the four classes identified in Table~\ref{tbl:multilab}. We use the last model in RQ3 but we use multi-label as target classification and walkthrough method as a research variable. We use 1-min live data in 2018 for 1 month-window and the time interval of data is randomly selected. 

\begin{figure}[h]
\centering
\includegraphics[width=.8\columnwidth]{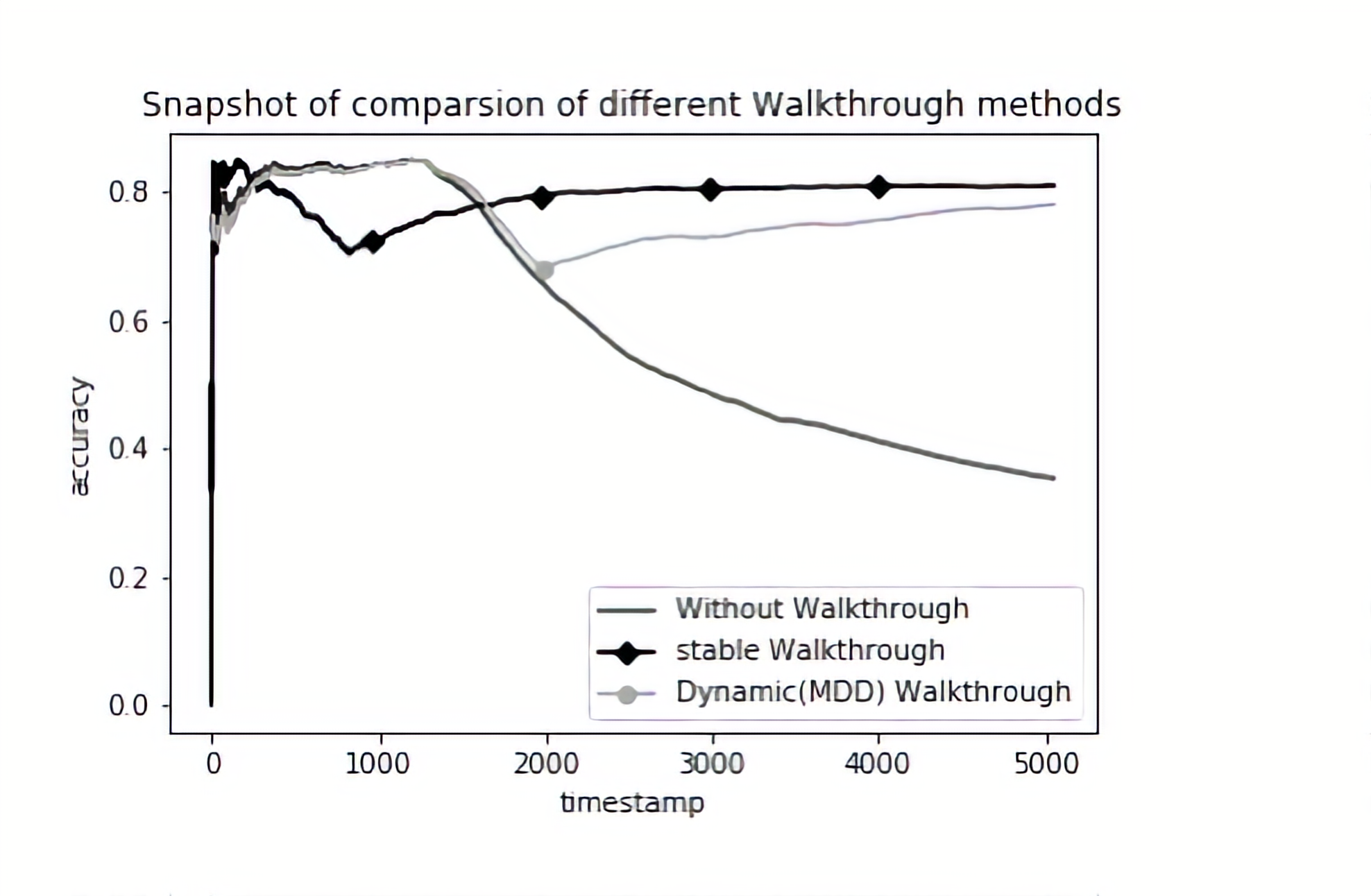}
\caption{Snapshot of comparison between different walkthrough methods}
\label{fig:walkthrough}
\end{figure}

Figure~\ref{fig:walkthrough} gives an overview of the comparison between the three different walkthrough methods we tested. 

For the first method, we train the deep learning model statically without walkthough, which means we will not retrain or update the model when the accuracy decays. We can observe how the accuracy drops significantly after roughly 1,500 predictions and reaches a value of less than 40\%, almost as low as random-guessing (25\%), in the end. 

When we use stable walkthrough to train our model, we will retrain at regular time periods. Our tests are based on trading period or a financial trading cycle of five days. On our data, this leads to four retrains (identified by rectangular points on the accuracy line). The first retrain point has obvious effects in improving accuracy and it starts to go up slightly before retraining. After four instances of retrain, the accuracy stabilizes around 80\%. 

When we use MDD-Dynamic Walkthrough method, we retrain the original model when the accuracy drops by more than 15\%. In our test, there is only one such instance (circle point at around 2,000 mark). The accuracy has apparent growth after this model adjustment. 

\begin{figure}[H]
\centering
\includegraphics[width=0.9\columnwidth]{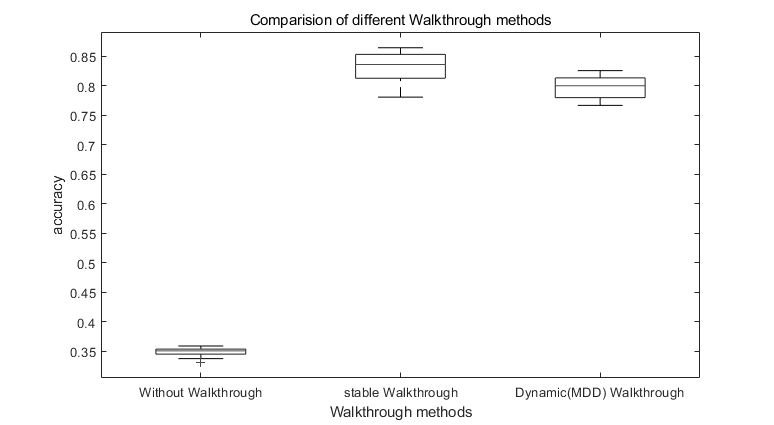}
\caption{Comparison of different Walkthrough methods}
\label{fig:batch}
\end{figure}

We also perform the experiment for 20 times to compare the different walkthrough methods in order to have a stronger statistical guarantee; results are shown in Figure \ref{fig:batch}, (Considering that repeated experiments cost significant computation power and time, we repeat the experiment 20 times to gather the results.) The results are in line with those discussed above, i.e., stable walkthrough is better than the other two methods. The results also show that, for dynamic (MDD) walkthrough method, only 2/3 retraining points occur in most experiments (90\%) while 2 experiments require 5 retraining points. But for stable walkthrough method, all experiments need 4 retraining points. Therefore, when considering retrain time is a factor, dynamic (MDD) walkthrough method is better because it needs less retraining in most cases. 

\textbf{The answer to RQ4}  is multifold. Firstly, in multi-label prediction with deep learning model, walkthrough training significantly improves the prediction accuracy. The reason is that, as discussed above, deep learning prediction models need to update itself. When the model is not fit for the new market conditions, then it must be updated to achieve accurate results. Secondly, stable walkthrough method is better than MDD-dynamic walkthrough method, unless retrain time is important. 

\section{Validity of findings}\label{sec:lim}
The model is based on trading system; we select historical data and collect live data for a long time span. The selection of data has no bias because historical trading contains all available transaction data and available currency pairs in cryptocurrency market. Moreover, the experiments are not affected by bull or bear market, policy impact and other factors. The experiments use an extensive data selection, including bull-market condition, bear-market condition, high-transaction-volume condition, low-transaction-volume condition etc. 

Quality of data is another important factor to discuss. As the data is collected live from Coinbase Pro, poor connection might affect the data (e.g., missing values). To mitigate this risk, we have compared the data collected from Coinbase Pro with other third party service providers to make sure the experiment have not been affected by inappropriate financial data. 

\section{Conclusions}\label{sec:con}
This paper analyzes a data-driven approach to predict mid-price movements in cryptocurrency markets, and covered a number of research questions en route regarding parameter settings, design of neural networks and universality of the model. The main finding of our work is the successful combination of an autoencoder and a walkthrough retraining method to overcome the decay in predictive power on live data due to non-stationary features on the order book. Prediction in high-frequency cryptocurrency markets is a challenging task because the environment contains noisy information and is highly unpredictable. We believe that our results can inform the design of higher level trading strategies and our networks architecture can be used as a feature to another estimator.


\bibliographystyle{agsm}
\bibliography{ref}

\end{document}